\documentclass[epj]{webofc}
\usepackage[varg]{txfonts}   

\wocname{EPJ Web of Conferences}

\woctitle{CONF12}

\usepackage{placeins}
\usepackage{tabularx}
\usepackage{amsmath}
\usepackage{wrapfig}
\usepackage{mathtools}

\begin{document}

\selectlanguage{english}
\title{Quarkonia at $T>0$ and lattice QCD}

\author{Alexander Rothkopf\inst{1}\fnsep\thanks{\email{rothkopf@thphys.uni-heidelberg.de}} }

\institute{Institute for Theoretical Physics, Universit\"at Heidelberg, Philosophenweg 12, D-69120 Germany}

\abstract{
We report here on recent progress in the determination of S-wave and P-wave heavy-quarkonium states at finite temperature. Our results are based on the combination of effective field theories with numerical lattice QCD simulations. These non-perturbative tools allow us to compute the heavy-quarkonium in-medium spectral functions, from which we in turn determine the melting temperatures of individual states and estimate phenomenologically relevant observables, such as the $\psi^\prime$ to J/$\psi$ ratio in heavy-ion collisions.}

\maketitle

\section{Introduction}

The bound states of a heavy quark and anti-quark ($c\bar{c},b\bar{b}$), christened heavy-quarkonium (charmonium, bottomonium), have matured into a high precision probe in the study of heavy-ion collisions. Their constituents are produced early on in the partonic stages of a collision and subsequently sample the full evolution of the quark-gluon plasma (QGP) created in the collision center. Both the experimental programs at RHIC and LHC have over the last decade accumulated an unprecedented amount of high-precision data on these states. Among the highlights from run1 at the LHC \cite{Andronic:2015wma} are the observation of relative excited states suppression in the bottomonium S-wave channel by the CMS collaboration, as well as the finding by the ALICE collaboration that the yields for the charmonium ground state $J/\psi$ are replenished at $\sqrt{s_{NN}}=2.76$TeV, compared to those measured at RHIC. The availability of such detailed measurements now urges theory more than ever to provide first-principle insight into the physics of heavy $Q\bar{Q}$ in heavy-ion collisions.

When we wish to start from first principles, naturally some idealization of the actual situation encountered in experiment is necessary. Here our main assumption is that of full kinetic equilibration of the heavy quarks with their surrounding, which is furthermore taken to be a thermal medium of quarks and gluons at a fixed temperature $T$. Since the temperatures encountered in current heavy-ion collision are not higher than $3-4\times T_{\rm PC}$ with $T_{\rm PC}=155\pm5$MeV \cite{Bazavov:2011nk,Borsanyi:2013bia} the chiral crossover temperature, we will use non-perturbative lattice QCD simulations to describe the  medium degrees of freedom.

In such a static setup, we may compute the thermal spectral functions of heavy quarkonium, which provide us with vital information about their in-medium behavior. At $T=0$ bound states appear as almost delta-like peaks, positioned at the mass of the state and well separated from the open heavy-flavor threshold, at which the spectral function rises to a continuum. Their binding energy, both in vacuum and in-medium, is defined from the distance between the peak position and this continuum. At finite temperature these peaks may shift and thermally broaden and will eventually dissolve into the continuum. Besides the position and width of peaked structures, their area plays an important role too, as it is linked with the decay of the corresponding in-medium state either into dileptons or light hadrons \cite{Bodwin:1994jh}. I.e. using the position, width and area of the in-medium spectra, we will set out to estimate what consequences in-medium modification will have on measured yields in heavy-ion collisions.

\section{Quarkonium in-medium spectral functions}

For theory, heavy-quarkonium represents a unique probe, due to the inherent separation of scales between the heavy-quark rest mass $m_Q$ and e.g.\ the temperature $T/m_Q\ll 1$ or the characteristic scale of QCD $\Lambda_{\rm QCD}/m_Q\ll 1$. In the presence of such a separation, we may use the small ratios as expansion parameters to systematically go over from the language of quantum chromodynamics (QCD) to a non-relativistic description of the heavy quarks in terms of effective field theories (EFT) \cite{Brambilla:2004jw}. We will hence compute the in-medium quarkonium spectral functions in the following using two complementary EFT approaches combined with lattice QCD simulations: The first and direct one is based on non-relativistic QCD (NRQCD) \cite{Thacker:1990bm,Lepage:1992tx}, which treats realistic heavy-quarkonium with a finite mass. However with the currently available simulation data its resolution is still limited. This line of study is also pursued by the FASTSUM collaboration in Refs.~\cite{Aarts:2010ek,Aarts:2011sm,Aarts:2014cda,Aarts:2013kaa}. The second approach is an indirect one, based on the EFT potential NRQCD (pNRQCD) (for a perturbative study see \cite{Brambilla:2008cx,Burnier:2007qm}), where one first computes the in-medium potential between the $Q\bar{Q}$ and in turn solves a Schr\"odinger equation for the in-medium spectral function. And while this approach is not limited in resolution, the underlying $T>0$ potential we use at the moment does not yet include finite velocity or spin dependent corrections.

Both approaches face the same challenge that at some stage they need to extract dynamical information in the form of spectral functions from lattice QCD simulations, which are carried out in an unphysical Euclidean time. The simulated correlation functions $D(\tau_i)=D_i$ are related to the spectral functions via a Laplace-type integral transform, which needs to be inverted
\begin{align}
D(\tau)=\int_{-2m_Q}^\infty d\omega e^{-\omega \tau} \rho(\omega),\quad D_i=\sum_{l} \Delta \omega_l e^{-\omega_l\tau_i} \rho_l,\quad i\in[1\ldots N_d],\;l\in[1\ldots N_\omega],\; N_\omega\gg N_d\label{Eq:InvProb}.
\end{align}
Since we need to discretize the spectrum at many more frequencies, than we have simulation data available and each $D_i$ comes with a finite error, the inversion is inherently ill-defined. I.e.\ a $\chi^2$-fit of the $\rho_l$'s would lead to an infinite number of degenerate solutions that all reproduce $D_i$ within uncertainties. One possibility to nevertheless give meaning to the inversion is to use Bayes theorem \cite{Jarrell:1996}, which provides a systematic prescription how to regularize the naive $\chi^2$ fit by incorporating additional knowledge (I) on the spectra, e.g. positive definiteness or smoothness conditions. This prior knowledge is usually encoded in a function $m_l$ called the default model, which represents the extremum of the regulator functional, the prior probability $P[\rho| I]$
\begin{align}
P[\rho|D,I]\propto \underbracket{P[D|\rho,I]}_{\chi^2\, \rm likelihood\,probability} \cdot \underbracket{P[\rho|I]}_{\rm prior\,probability}, \qquad \left| \frac{\delta P[\rho|D,I]}{\delta \rho} \right|_{\rho=\rho^{\rm Bayes}}=0.
\end{align}
Once the regulator and $m$ is specified, one ends up with a numerical optimization problem to determine the most probable spectrum given simulation data and prior information. In the following we will use two different implementation on the market, the well known Maximum Entropy Method (MEM) \cite{Asakawa:2000tr} and a more recent Bayesian Reconstruction (BR) method \cite{Burnier:2013nla}. They differ both in the form of the regulator and the implementation of the numerical search for $\rho^{\rm Bayes}$. Over the last two years we have gained a much better understanding of the very different systematic uncertainties of the two methods: the MEM has been found to be prone to over-smoothing, in particular if only a relatively small number of datapoints is available, while the BR method can introduce ringing artifacts that may mimic peak features not actually present in the simulation data. Note that as long as we are not in the "Bayesian continuum limit", i.e. $N_d\to\infty$ and $\Delta D/D\to0$ two different implementation of the Bayesian strategy may provide different answers and only in that limit will agree. Unfortunately due to the strong loss of information due to the integral Kernel in \eqref{Eq:InvProb} approaching this limit is exponentially hard. These challenges will be apparent in the direct analysis of spectral functions using NRQCD on the lattice in the following subsection. 

\subsection{Direct determination based on lattice NRQCD}

Instead of simulating the light and heavy d.o.f. on the same space-time lattice, which would require too fine of a lattice spacing, we here \cite{Kim:2014iga} use the separation of scales $T/m_Q\ll 1$, $\lambda_{\rm QCD}/m_Q\ll 1$ to separate the physics of the thermal medium from that of the heavy quarks in the following way. In this lattice variant of the EFT NRQCD, the light quarks and gluons are treated with a standard lattice QCD simulation at $T>0$, while the heavy quarks are described by non-relativistic Pauli-spinors propagating in the background of the thermal fields. At this point no modelling is involved, as NRQCD is systematically derived from the QCD Lagrangian by an expansion in $1/m_Q a$ with $a$ being the lattice spacing. Current implementations include up to fourth order corrections, corresponding to ${\cal O}(v^4)$ in the language of continuum NRQCD. For the medium we utilize state-of-the-art lattices by the HotQCD collaboration \cite{Bazavov:2014pvz} on $48^3\times12$ grids featuring dynamical $u,d$ and $s$ quarks based on the HISQ formulation with almost physical pion mass $m_\pi=161$MeV. These simulations span temperatures $T\in[140\ldots407]$MeV with the inverse of the naive expansion parameter of lattice NRQCD for Bottomonium taking on the values $m_b a\in[2.759\ldots0.954]$. The results shown here are preliminary as they do only include a subset of the available lattice simulations.

\begin{figure}[t!]
\centering
\includegraphics[scale=0.6]{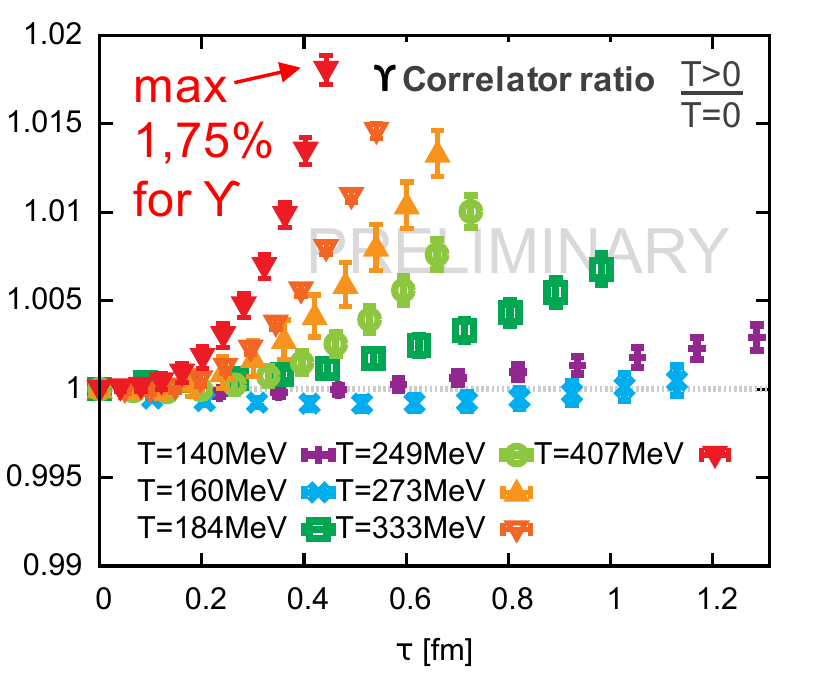}
\includegraphics[scale=0.61]{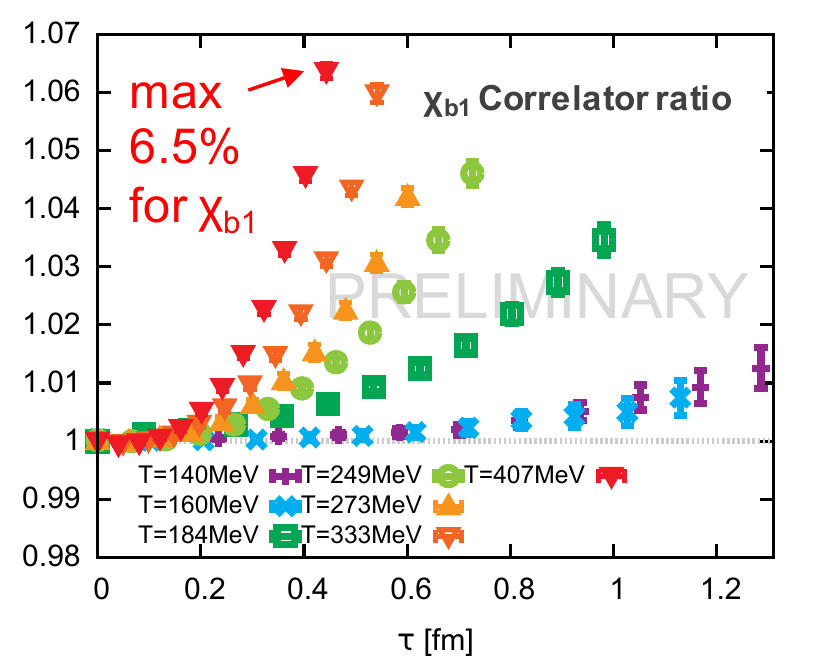}\vspace{-0.25cm}
\caption{Finite temperature to vacuum correlator ratios from lattice NRQCD for the Upsilon (left) and $\chi_b$ (right) channel.}\label{Fig:NRQCDCorrRatios}
\end{figure}

One then computes the propagation of a single heavy quark in the medium background, and combines two of these propagators into a heavy quarkonium correlator. Projected to a definite quantum number by the insertion of appropriate vertex operators one arrives at a heavy quarkonium correlation function in Euclidean time. This quantity can already tell us about overall in-medium modification if we divide its value at $T>0$ by that in vacuum, as shown in Fig.~\ref{Fig:NRQCDCorrRatios} for Upsilon (left) and $\chi_b$ (right).

If the ratio is unity within statistical errors, no in-medium modification is present. Around $T_C$ we find only very mild deviations from one, albeit there is a hint for a non-monotonic behavior present. At higher temperatures a characteristic upward bending occurs but even at the highest $T=407$MeV the maximum departure from unity is $1.75\%$ for Upsilon and $6.5\%$ for $\chi_b$. The differences between the two cases however are a first sign that in-medium modification is actually hierarchically ordered with the vacuum binding energy of the ground state of a particular channel. 

\begin{figure}[t!]
\centering
\includegraphics[scale=0.5]{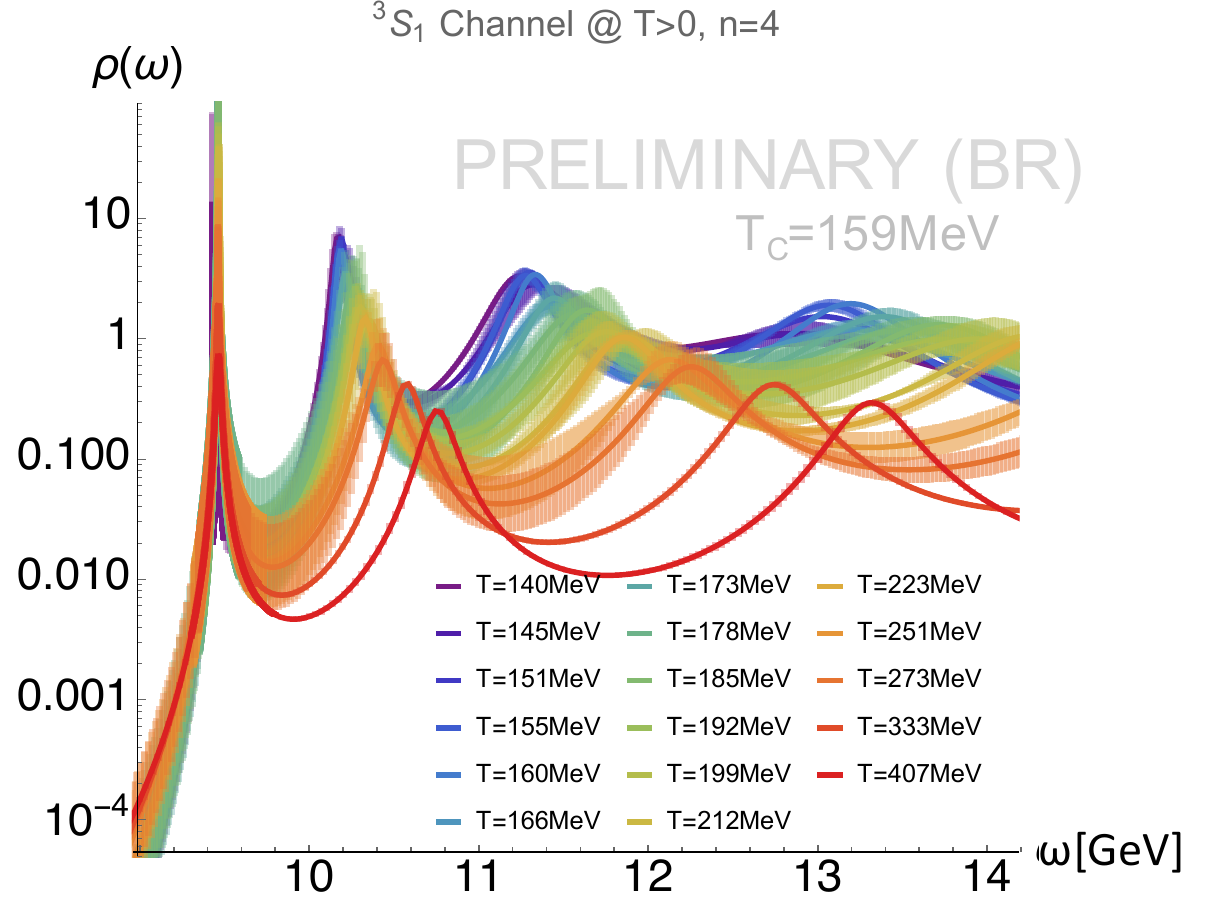}
\includegraphics[scale=0.5]{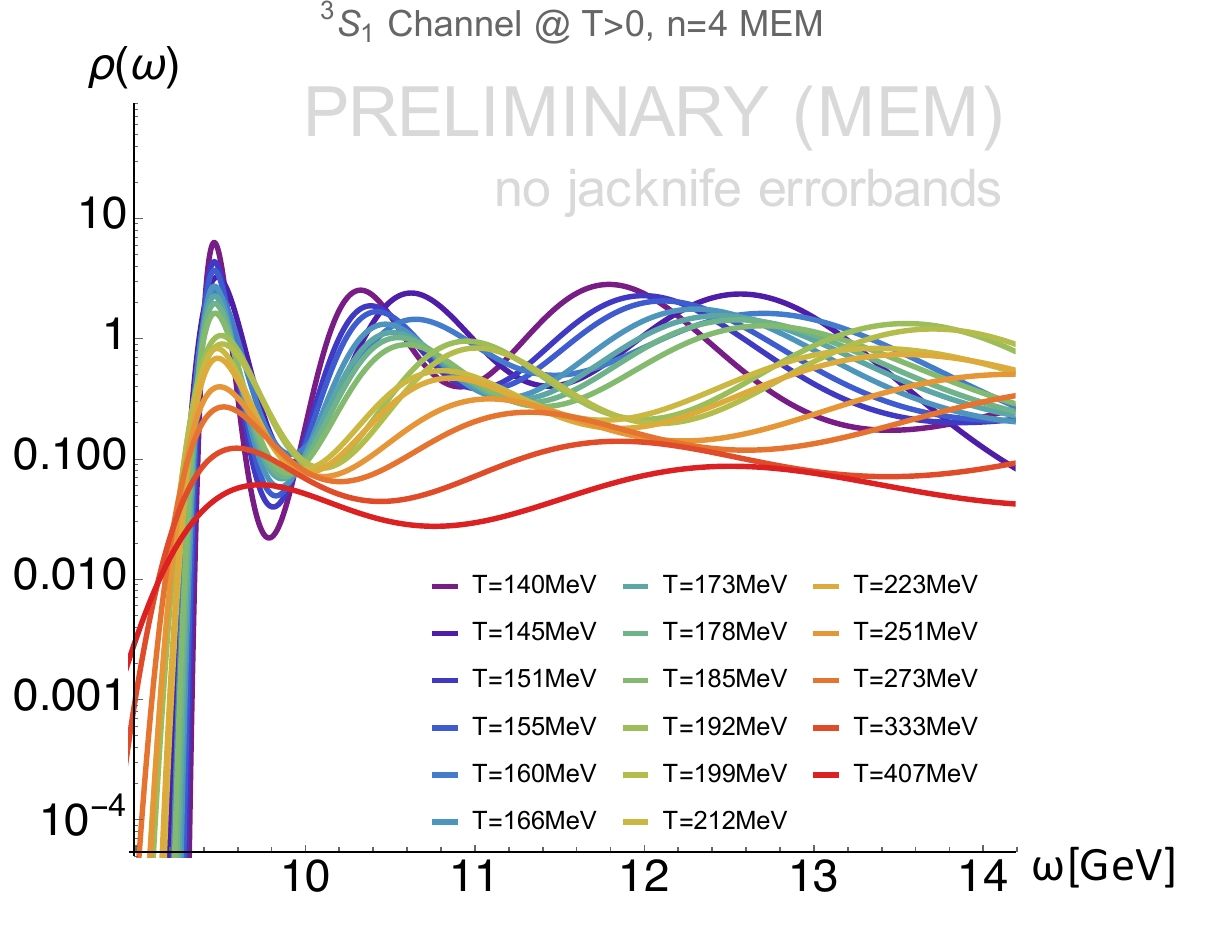}\vspace{-0.25cm}
\caption{Finite temperature Upsilon spectra from a direct reconstruction from lattice NRQCD correlation functions between $T\in[140\ldots407]$MeV. (left) Reconstructions based on the BR method show a well defined ground state feature at all temperatures, while (right) the same simulation data reconstructed with the MEM shows only washed out features above $T\approx 333$MeV. Due to the different systematics of the two methods they bracket the actual disappearance of the bound state peak from above (BR) and below (MEM).}\label{Fig:NRQCDSpecsUpsilon}
\end{figure}

From the in-medium correlators we may now reconstruct the spectral functions using both the BR method and the MEM. We use a flat default model $m$ in the following and discretize frequencies with $N_\omega=3000$ leading us to the results for the Upsilon channel shown in Fig.~\ref{Fig:NRQCDSpecsUpsilon}. Note that due to the small number of simulation datapoints $N_d=12$ available only the ground state peak structure is captured robustly, all other higher lying features must be attributed to numerical artifacts.

\begin{figure}[t!]
\centering
\includegraphics[scale=0.5]{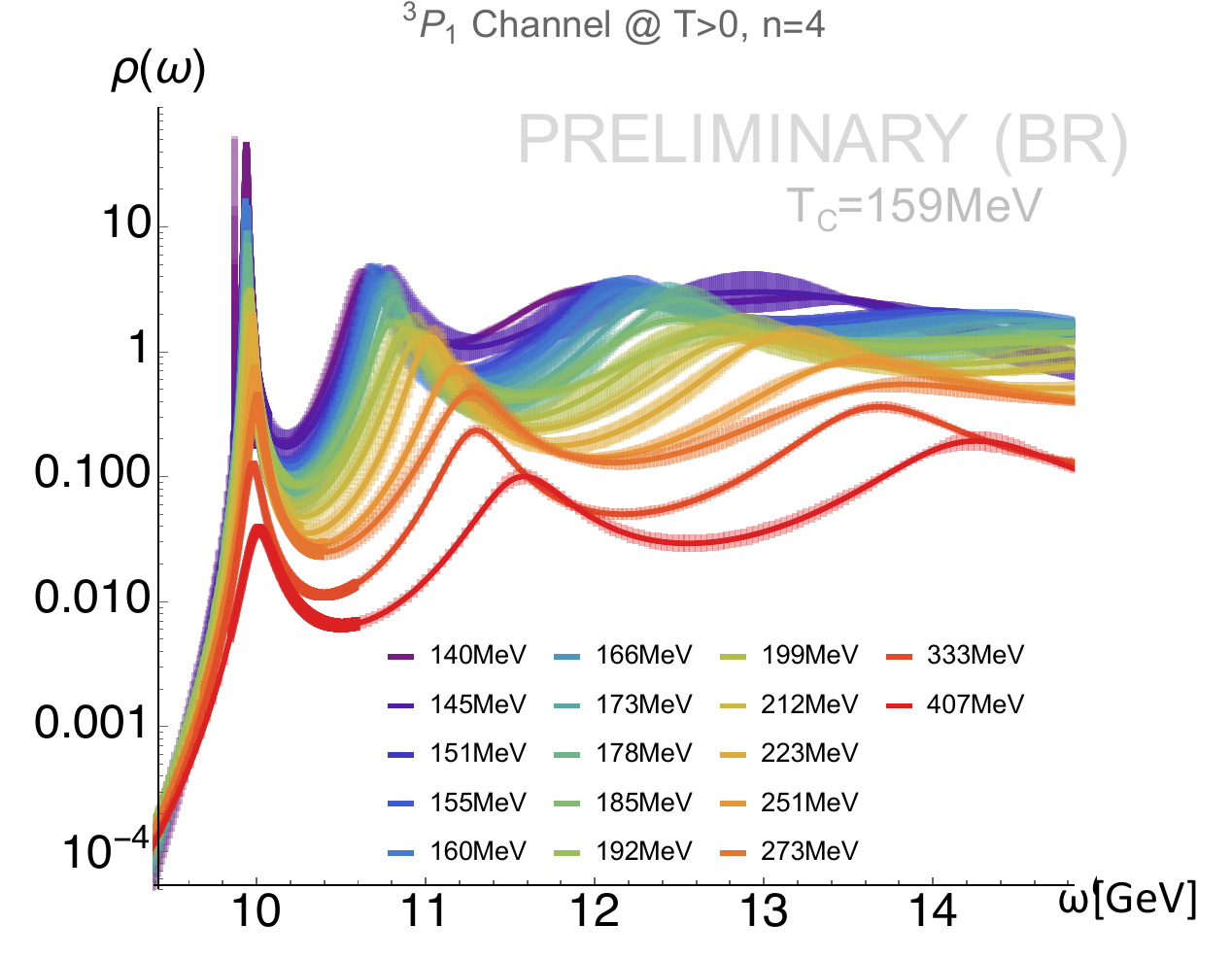}
\includegraphics[scale=0.5]{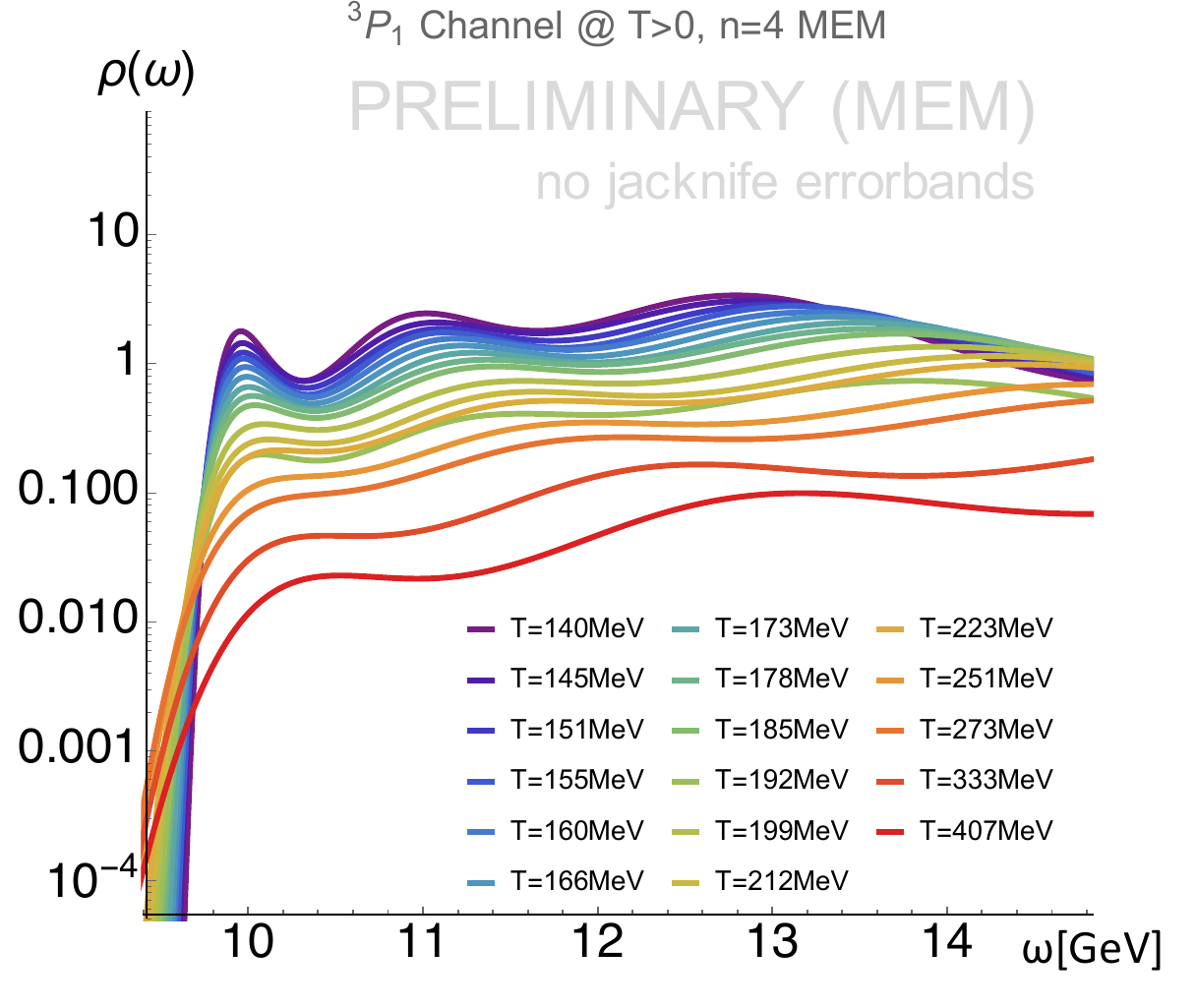}\vspace{-0.25cm}
\caption{Finite temperature $\chi_b$ spectra from a direct reconstruction from lattice NRQCD correlation functions between $T\in[140\ldots407]$MeV. (left) Reconstructions based on the BR method show features at the position close to the vacuum ground state at all temperatures, while (right) the MEM shows only washed out features above $T\approx 210$MeV. As described in the text, we can identify that the apparent ground state feature at $T=407$MeV in the BR method is a numerical artifact and thus both methods see the disappearance of $\chi_b$ below $T=407$MeV. Due to the different systematics of the two methods they however bracket the actual disappearance of the bound state peak from above (BR) and below (MEM). }\label{Fig:NRQCDSpecsChib}
\end{figure}

We find that the BR method shows a well pronounced lowest lying structure at all temperatures up to $T=407$MeV, i.e. the amplitude of the lowest peak is at least a factor 5 larger than the amplitude of the neighboring artifacts. The MEM on the other hand shows only washed out features above $T\approx 333$MeV. Similar MEM results have been obtained by the FASTSUM collaboration in \cite{Aarts:2014cda}. 

To interpret our results we need to consider the systematic uncertainties of the two different methods. The possibility for over-smoothing in the MEM and ringing in the BR method. I.e. one is bracketing the actual temperature for the disappearance of a state from above and below with the two methods. In the Bayesian continuum limit these temperatures would become the same.

Now let us continue to the P-wave states given in Fig.~\ref{Fig:NRQCDSpecsChib}. The BR method results shown on the left, at first sight, seem to indicate that again up to $T=407$MeV a well defined lowest lying peak is present, while the MEM, consistent with previous FASTSUM computations shows only washed out features above $T\approx210$MeV. 

A simple inspection by eye is however not enough in case of the BR method. Since the amplitude of the lowest peak at $T=407$MeV is already lower than that of the neighboring numerical artifact at higher frequencies, we need a more systematic way to determine whether that feature really represents a sign of a remnant bound state. Our strategy to do so is to compare with reconstructed non-interacting spectral functions. The actual free spectrum is known to be a continuous functions devoid of peaks. However if such spectrum is reconstructed from a small number of correlator points, akin to Gibbs ringing in the inverse Fourier transform we expect the BR method to also introduce wiggly behavior. \begin{wrapfigure}{r}{0.45\textwidth}\vspace{-0.3cm}
  \begin{center}
   \includegraphics[scale=0.475]{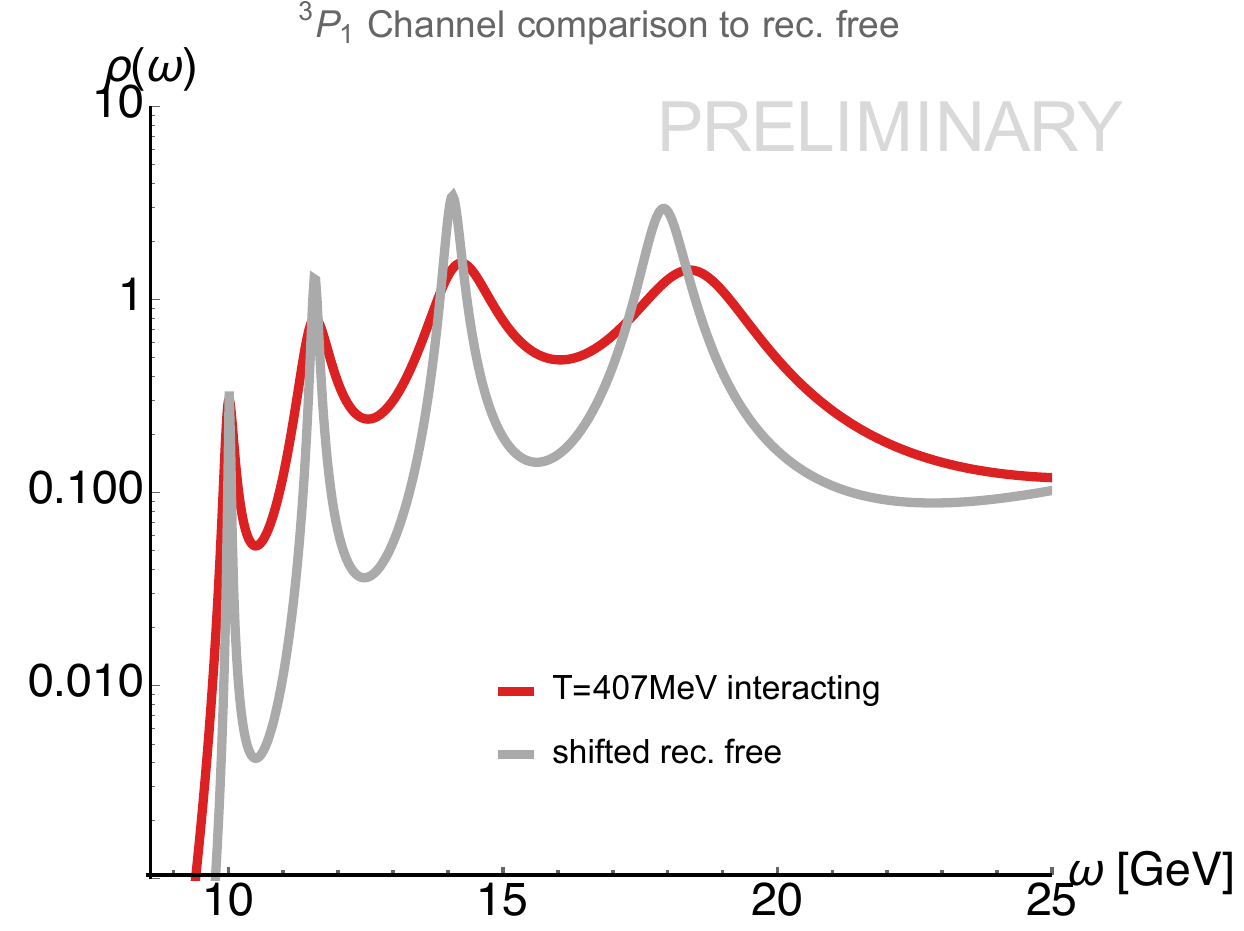}
  \end{center}\vspace{-0.6cm}
 \caption{The reconstructed interacting (red) and appropriately shifted reconstructed free (gray) spectral function at $T=407$MeV.}\label{Fig:NRQCDSpecsChibFree}\vspace{-0.5cm}
\end{wrapfigure} And indeed as shown in Fig.~\ref{Fig:NRQCDSpecsChibFree} the gray curve corresponding to the reconstructed free spectrum shows sizable ringing artifacts. In particular their amplitude is at least as strong as that of the apparent bound state signal in the interacting reconstructed spectrum. We conclude that the BR method does not give any indication of a bound state remnant at $T=407$MeV and note again that by using the two different methods, BR and MEM, we are thus bracketing the actual disappearance of the $\chi_b$ bound state from above and below.

While this direct approach to in-medium quarkonium based on lattice NRQCD is promising the available simulation data needs to be improved to access also the physics of e.g. excited states in this framework. In the spirit of reaching the Bayesian continuum limit, we are thus currently working on increasing the statistics, while e.g.\ the FASTSUM collaboration is generating ensembles with a larger number of points $N_d$ along Euclidean time direction. 

\subsection{Indirect determination based on pNRQCD}

\begin{figure}[t!]
\centering\vspace{-0.6cm}
\includegraphics[scale=0.27]{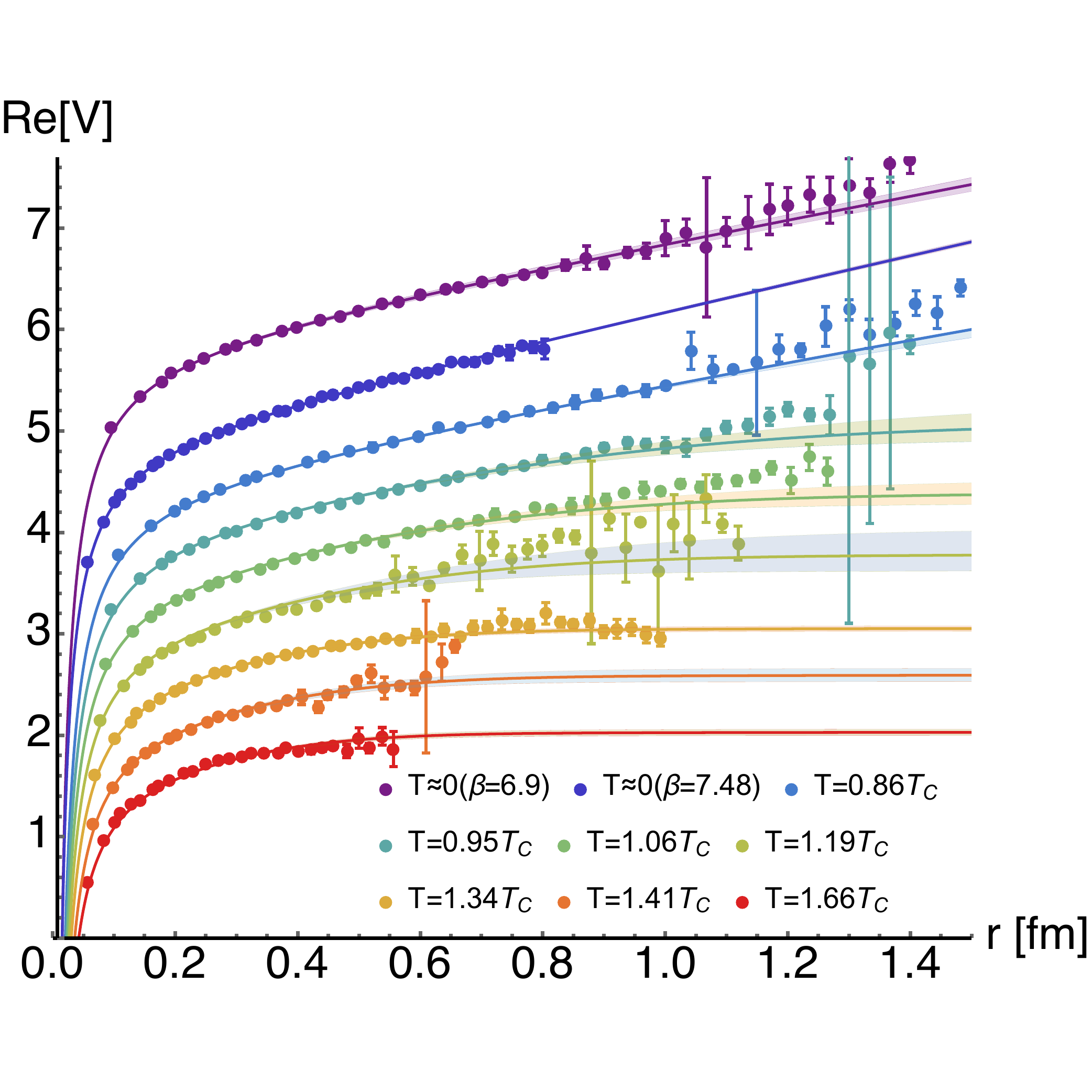}\hspace{0.4cm}
\includegraphics[scale=0.27]{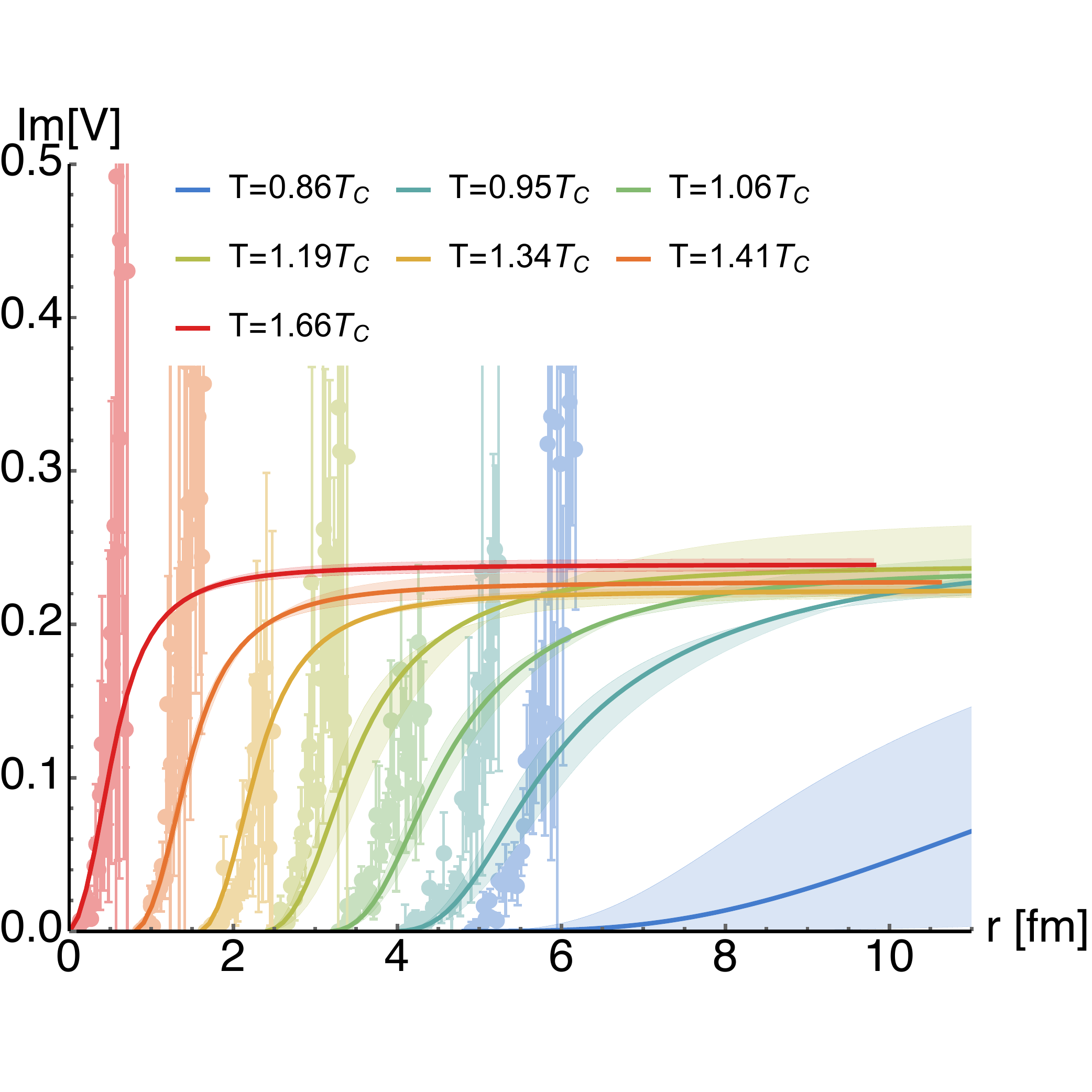}\vspace{-0.4cm}
\caption{ (left) Real- and (right) imaginary part of the static in-medium heavy-quark potential. The values of ${\rm Re}[V^{\rm QCD}]$ are shifted in y and those of ${\rm Im}[V^{\rm QCD}]$ in x for better readability. The filled symbols correspond to the actual values obtained from full lattice QCD and the errorbars arise from a ten bin jackknife. The solid lines correspond to a analytic parametrization of the in-medium potential based on a generalized Gauss law ansatz, in which all in-medium effects are encoded in a single temperature dependent parameter $m_D$, which we identify with the Debye mass.}\label{Fig:QQbarPotential}
\end{figure}

A complementary but indirect approach to in-medium heavy quarkonium spectral functions is based on the effective field theory pNRQCD. In it the $Q\bar{Q}$ two-body system is described in terms of non-relativistic color singlet and color-octet wave-functions. The correlation functions of such wavefunctions evolve under a Schr\"odinger like equation of motion featuring a potential $V^{\rm QCD}(r)$ that encodes the interactions between the constituents and the environment. As the EFT can be systematically derived from QCD this potential consists of a lowest order static contribution and may be amended by velocity and spin dependent corrections. Depending on the degree of scale separation also non-potential effects may play a role but are not further considered here. The values of $V^{\rm QCD}$ are obtained from the underlying microscopic field theory by the process of matching, where two correlation functions of equal physical content, one in pNRQCD, one in QCD are set equal at a certain scale. It has been shown that the in-medium potential can be related to a real-time quantity in thermal QCD, the rectangular Wilson loop
\begin{align}
V^{\rm QCD}(r)=\lim_{t\to\infty} \frac{i\partial_t W_\square(t,r)}{W_\square(t,r)}, \quad W_\square(t,r)=\Big\langle {\rm Tr} \Big( {\rm exp}\Big[-ig\int_\square dx^\mu A_\mu^aT^a\Big] \Big) \Big\rangle\label{Eq:VRealTimeDef}.
\end{align}
Evaluating this definition in resummed perturbation theory at high temperature \cite{Laine:2006ns,Beraudo:2007ky} revealed that it actually can take on complex values, i.e. it may contain an imaginary part, related to scattering of light medium d.o.f. off the color string spanning between the $Q\bar{Q}$. The fact that lattice QCD simulations are carried out in unphysical Euclidean time prohibits us from directly evaluating \eqref{Eq:VRealTimeDef}. On the other hand the concept of spectral function $\rho_\square$ (not the quarkonium spectral function but that of the Wilson loop) allows us to connect the Minkowski domain, where the potential is defined with the imaginary time domain in which numerical simulations are carried out \cite{Rothkopf:2009pk,Rothkopf:2011db}
\begin{align}
\nonumber W_\square(\tau,r)\hspace{-0.1cm}=\hspace{-0.15cm}\int d\omega e^{-\omega \tau} \rho_\square(\omega,r)\,
&\hspace{-0.1cm}\leftrightarrow\hspace{-0.17cm}\, \int d\omega e^{-i\omega t} \rho_\square(\omega,r)\hspace{-0.1cm}=\hspace{-0.1cm}W_\square(t,r),\quad V^{\rm QCD}(r)=\lim_{t\to\infty} \frac{ \int d\omega \omega e^{-i\omega t} \rho_\square(\omega,r)}{\int d\omega\, e^{-i\omega t} \rho_\square(\omega,r)}. \label{Eq:PotSpec}
\end{align}
Based on a general argument using the symmetries of the real-time Wilson loop we have shown \cite{Burnier:2012az} that if a well defined lowest lying spectral feature exists in $\rho_\square$ a potential can be meaningfully defined. The values of ${\rm Re}[V^{\rm QCD}]$ and ${\rm Im}[V^{\rm QCD}]$ are furthermore related to the position and width of that lowest lying spectral feature. And since the Wilson loop spectral function is actually much simpler in structure than the quarkonium spectral function we are able to reconstruct it already with currently available simulation data (Note that in order to avoid cusp divergencies, in practice we compute the Wilson line correlator in Coulomb gauge instead of Wilson loops).  Here we use \cite{Burnier:2014ssa} lattice QCD simulations by the HotQCD collaboration with dynamical $u,d$ and $s$ quarks based on the asqtad formulation with heavier than physical pion mass $m_\pi\approx300$MeV. The corresponding outcome is shown as filled points in Fig.~\ref{Fig:QQbarPotential} for  ${\rm Re}[V^{\rm QCD}]$ on the left and ${\rm Im}[V^{\rm QCD}]$  on the right.

The real part is shifted by hand in y for better readability and shows a clear, as well as smooth transition from a confining behavior close to $T=0$ to a screened form above $T=T_C$. In the left panel we show as light colored points the values obtained for the imaginary part (shifted in x), which is finite above $T_C$. The collection of discrete points obtained so far however is not enough to compute the heavy quarkonium spectral function, to this end an analytic parametrization is required. 

Recently we derived such an analytic parametrization \cite{Burnier:2015nsa} starting from the realization that at $T\approx 0$ the values of ${\rm Re}[V^{\rm QCD}]$ we obtain on the lattice are very well described by a simple Cornell ansatz $V^{\rm QCD}_{\rm T=0}(r)=V_{\rm Cornell}(r)=-\alpha_S/r+\sigma r+c$. $\alpha_S$ denotes the strong coupling, $\sigma$ the string tension and $c$ is an arbitrary scale dependent shift. These three parameters in the following remain temperature independent and are determined via a fit to the $T\approx0$ datapoints on the left of Fig.~\ref{Fig:QQbarPotential}. The corresponding best fit is given as the solid dark violet and dark blue line. Similar to the Gauss law for the Coulombic part of the Cornell potential one can establish a generalized Gauss law to also describe the linearly rising part. 

In-medium effects are introduced in this setup by using the in-medium permittivity of a weakly coupled gas of quarks and gluons, taken from hard-thermal loop resummed perturbation theory. The idea is that the non-perturbative effects of inter-quark binding and confinement are encoded in the Cornell form of the $T=0$ potential and that the HTL contributions to the medium will as a first step provide us with the most relevant $T>0$ modifications. Combining this complex valued permittivity with the generalized Gauss law, we then obtain expressions for an in-medium modified real- and imaginary part arising from the Coulombic part of the $T=0$ potential but in addition also contributions to the real- and imaginary part from the string part of $V_{\rm Cornell}(r)$.

Since temperature enters the complex HTL permittivity only via a single temperature dependent parameter, i.e. the Debye mass $m_D(T)$ all modification of both ${\rm Re}[V^{\rm QCD}]$ and ${\rm Im}[V^{\rm QCD}]$ are determined by this single quantity. In turn, if we fix the value of $m_D$ using the robustly reconstructed ${\rm Re}[V^{\rm QCD}]$ our ansatz provides a postdiction of ${\rm Im}[V^{\rm QCD}]$.

\begin{wrapfigure}{r}{0.45\textwidth}\vspace{-0.6cm}
  \begin{center}
   \includegraphics[scale=0.5]{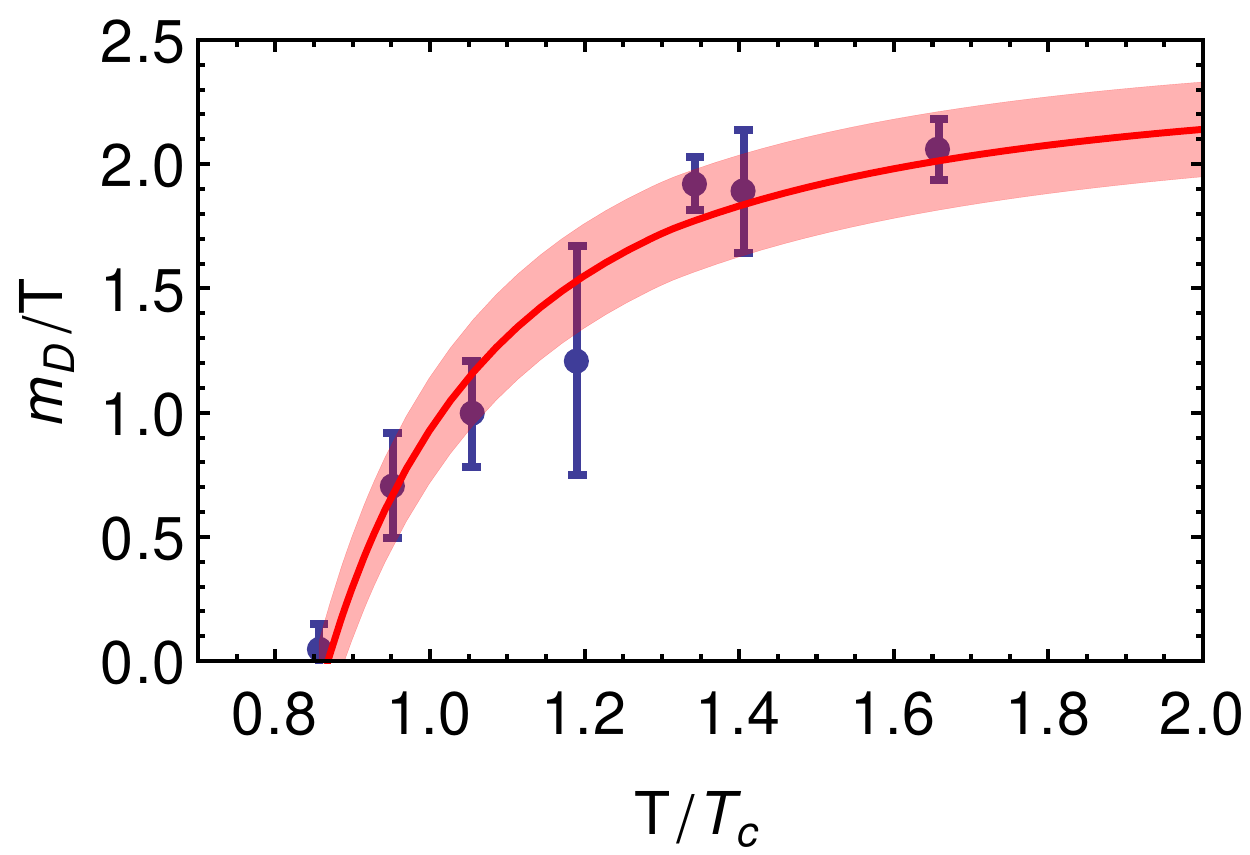}
  \end{center}\vspace{-.6cm}
 \caption{The best fit values for $m_D/T$ according to the lattice extracted ${\rm Re}[V^{\rm QCD}]$. The red solid line corresponds to a fit that smoothly goes over into the HTL perturbative form at high temperature}\label{Fig:mD}\vspace{-0.5cm}
\end{wrapfigure}While our ansatz contains several assumptions it manages to retrace the values of ${\rm Re}[V^{\rm QCD}]$ extracted on the lattice excellently, when tuning $m_D$ \cite{Burnier:2015tda}, as seen by the lower seven solid lines on the left of Fig.~\ref{Fig:QQbarPotential}. The corresponding values of $m_D/T$ are shown in Fig.~\ref{Fig:mD}, and are consistent with zero below $T_C$ while quickly switching to finite values above. The red solid line shows a fit, which approaches the perturbative HTL value of $m_D$ at very high $T\gg T_C$. 

To compute realistic heavy quarkonium spectra of course we would have to use continuum extrapolated values for the potential. As these are not yet available but still work in progress, we will use the following strategy. At $T=0$ we fit the values of the Cornell potential parameters so as to reproduce the spectrum of bottomonium bound states listed in the PDG. The in-medium modification is then imprinted via the Gauss-law parametrization and the continuum corrected values of $m_D$ obtained on the lattice.

\begin{figure}[t!]
\centering\vspace{-0.2cm}
\includegraphics[scale=0.27]{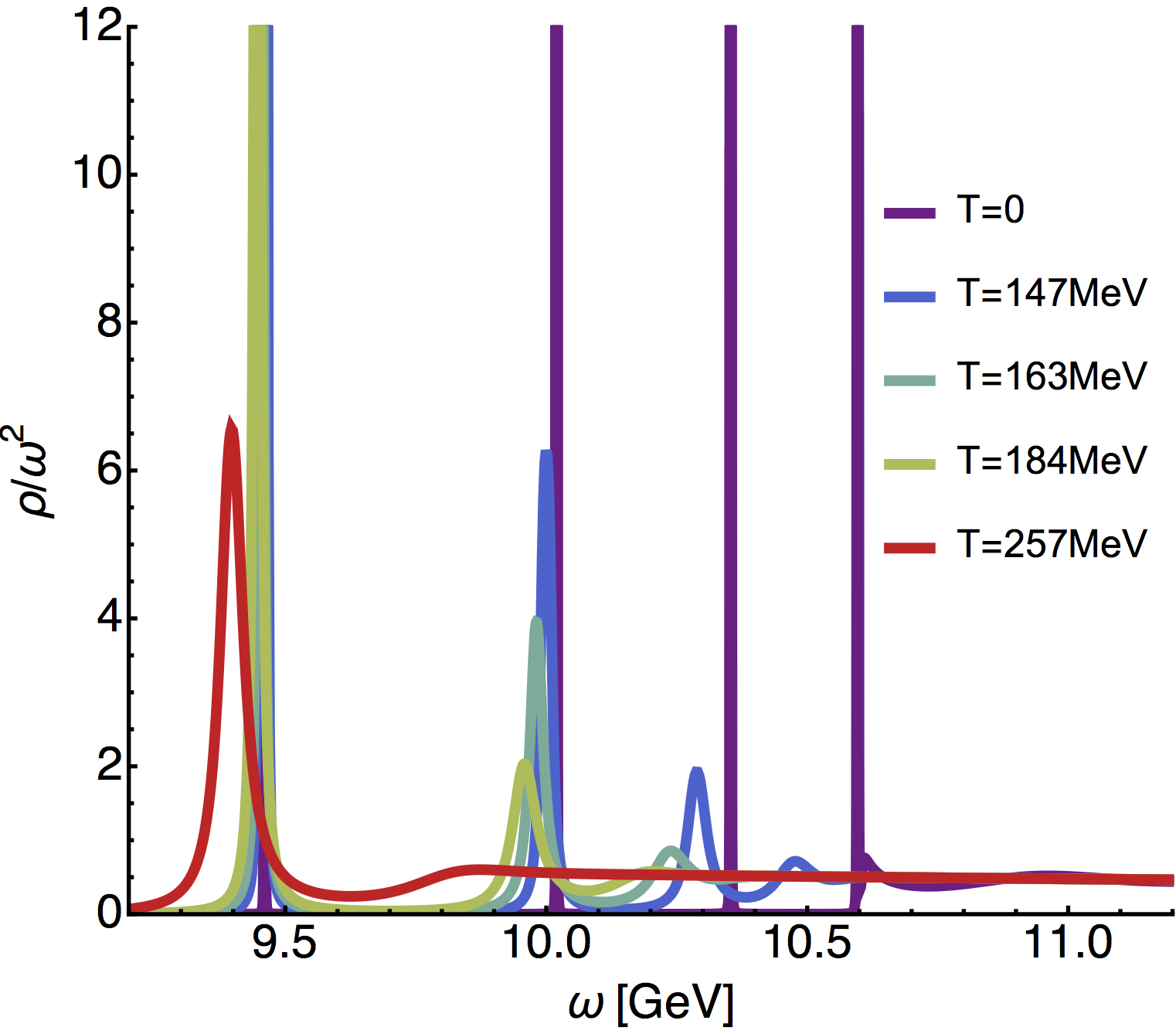}\hspace{0.3cm}
\includegraphics[scale=0.27]{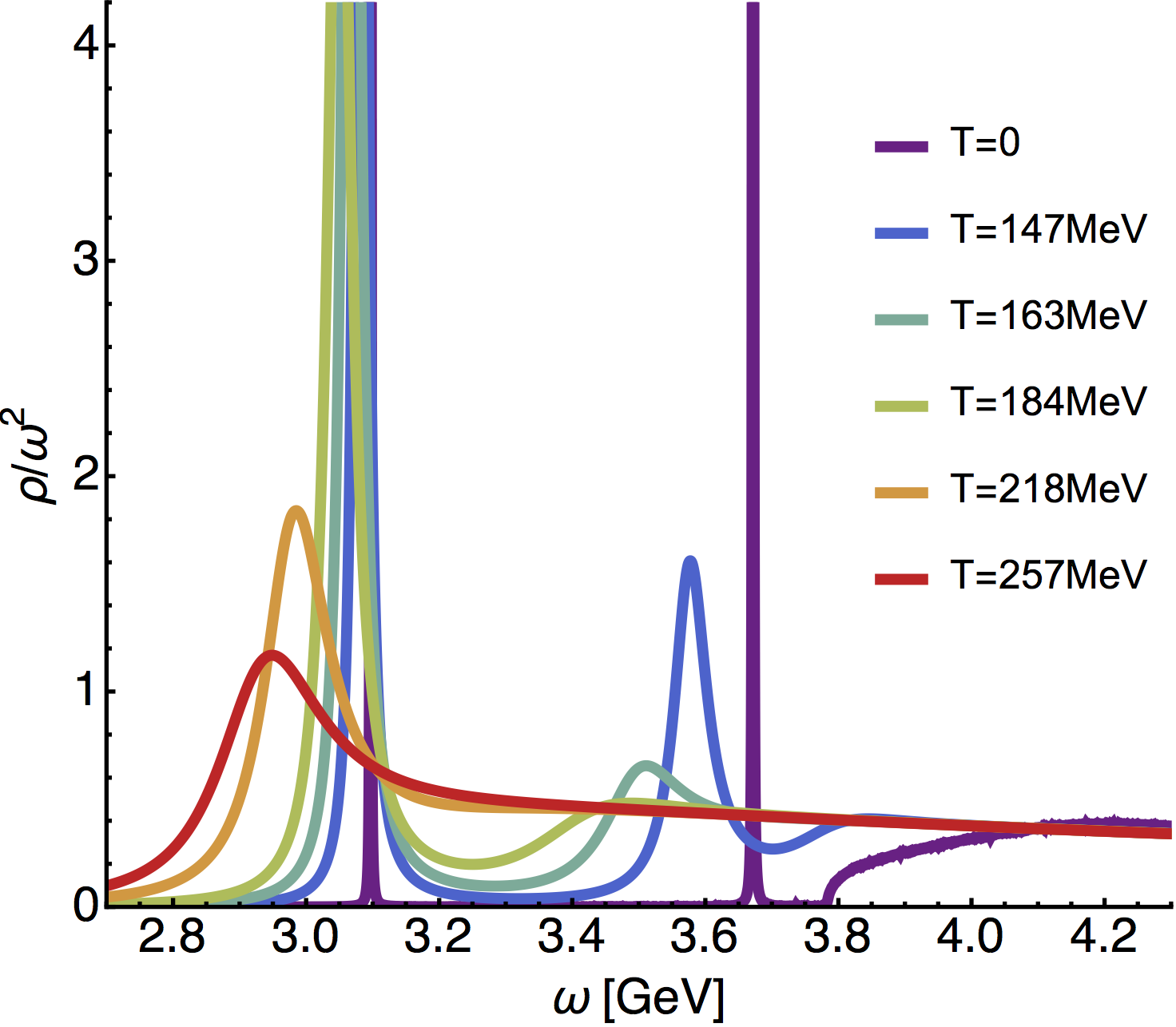}\vspace{-0.25cm}
\caption{In-medium S-wave bottomonium (left) and charmonium (right) spectra around the phase transition temperature. Note the hierarchical broadening and shifting to lower masses of individual state with respect to the vacuum binding energy. At the same time the continuum also moves to lower energies with increasing T.}\label{Fig:SwaveFromPot}
\end{figure}

\begin{figure}[t!]
\centering\vspace{-0.2cm}
\includegraphics[scale=0.27]{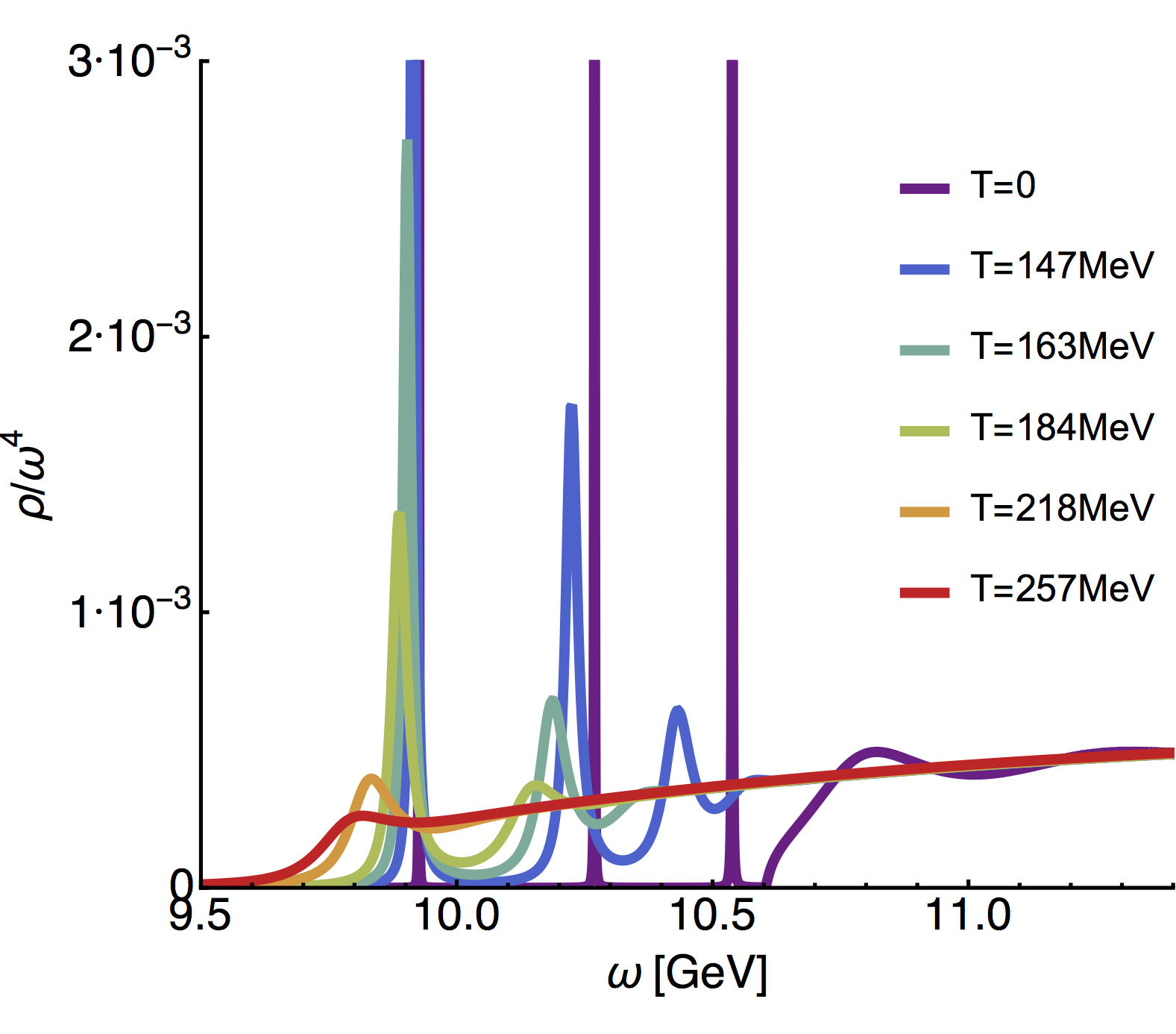}\hspace{0.3cm}
\includegraphics[scale=0.27]{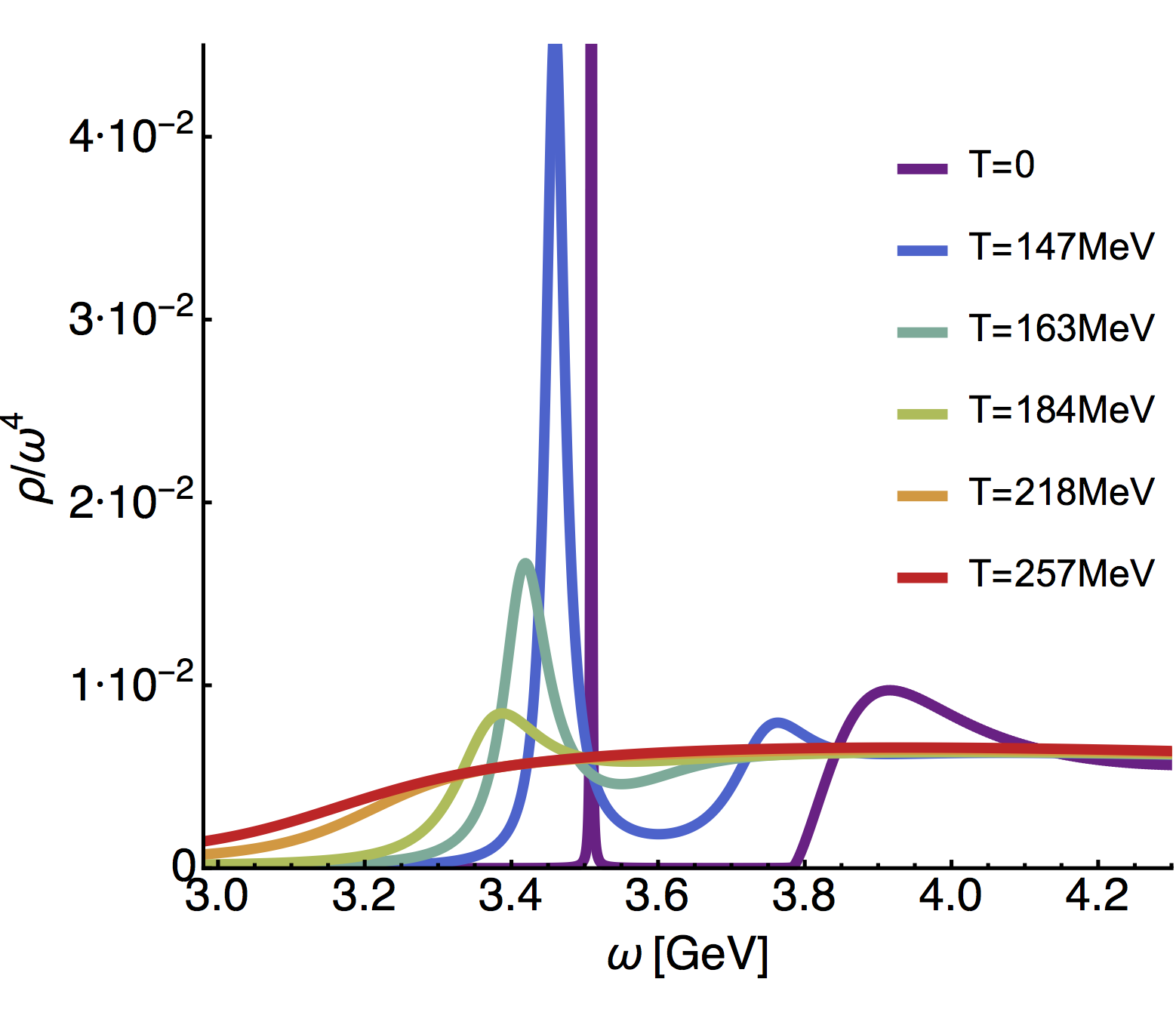}\vspace{-0.25cm}
\caption{In-medium P-wave bottomonium (left) and charmonium (right) spectra around the phase transition temperature. Note the hierarchical broadening and shifting to lower masses of individual state with respect to the vacuum binding energy. At the same time the continuum also moves to lower energies with increasing T.}\label{Fig:PwaveFromPot}
\end{figure}

With all these ingredients set, we can now compute the in-medium spectral functions by solving the Schr\"odinger equation for the so called forward correlator $D^>(t,r)$, which is in essence the unequal time correlator of singlet wavefunctions \cite{Burnier:2007qm}. Note that this is not the same as solving the Schr\"odinger equation for an individual realization of the wavefunction. The sought after quarkonium spectrum is related to this correlator via a Fourier transform and an appropriate limiting procedure $\lim_{r\to0}\int dt e^{-i\omega t} D(t,r)=\rho(\omega)$. The results for the S-wave channel \cite{Burnier:2015tda} are shown in Fig.~\ref{Fig:SwaveFromPot} and for the P-wave \cite{Burnier:2016kqm} in Fig.~\ref{Fig:PwaveFromPot}, the left panel contains the spectra for bottomonium, the right for charmonium respectively.


We observe characteristic changes in all the channels, i.e.\ a broadening and shifting of individual peaks to lower masses as temperature is increased. The strength of the modification is clearly ordered according to the vacuum binding energy of the individual states. Besides the bound state features also the position of the heavy-flavor threshold moves to lower energies, since the real-part of the potential becomes more and more weakened as $T$ rises. Eventually this continuum engulfs the bound state remnants and they disappear. This however is a gradual process and one needs to set a quantitative criterion at which $T$ we declare a state melted. With the information from the spectra available we may use the popular choice of $\Gamma(T)=E_{\rm bind}(T)$ (see e.g.\ \cite{Laine:2006ns}), i.e.\ the point at which the in-medium width equals its in-medium binding energy. In a naive wavefunction picture it correspond a dampening by a factor $1/e$ after one oscillation. The corresponding melting temperatures we obtain for the S-wave and P-wave are given below 
\begin{table}[h!]\centering
 \begin{tabular}{|c|c|c||c|c|c|c|}\hline
states& $ J/\psi(1S)$& $\psi'(2S)$ & $\Upsilon (1S) $ &$\Upsilon (2S) $ &$\Upsilon (3S) $ &$\Upsilon (4S) $ \\ \hline\hline
 $T_{\rm melt}^{\Gamma=E_{\rm bind}}/T_C $ &$1.37_{-0.07}^{+0.08}$& $<0.95$  & $2.66_{-0.14}^{+0.49}$ & $1.25_{-0.05}^{+0.17}$ & $ 1.01_{-0.03}^{+0.03}$ & $<0.95$\\   \hline
 \end{tabular}\vspace{-0.5cm}
 \end{table}
\begin{table}[h!]\centering \begin{tabular}{|c|c|c||c|c|c|}\hline
states& $ \chi_c(1P)$& $\chi_c(2P)$ & $\chi_b (1P) $ &$\chi_b (2P) $ &$\chi_b (3P) $ \\ \hline\hline
 $T_{\rm melt}^{\Gamma=E_{\rm bind}}/T_C $ &$1.04(3) $& $<0.95$  & $1.41(6)$ & $1.06(3)$ & $0.98(2)$ \\   \hline
 \end{tabular}
 \end{table}

Melting temperatures are interesting in their own right, however they are not directly measurable in experiment. Therefore in the next section we wish to discuss our proposal how to proceed towards extracting phenomenologically relevant observables, such as the $\psi^\prime$ to $J/\psi$ ratio from the in-medium spectra.

\section{In-medium quarkonium phenomenology from spectral functions}

To connect to experimental results, we need to keep two caveats in mind. First, one does not measure in-medium dilepton emission of heavy quarkonium in experiment but instead the decay of vacuum states long after the QGP has ceased to exist. Therefore comparing the in-medium spectra directly to measured spectra is not admissible. On the other hand our assumption of full kinetic equilibration is, if at all, only applicable to charmonium at low $p_T$ and mid-rapidity. Therefore in the following we concentrate on the lighter of the two flavors.

Then we may ask what are meaningful observables for charmonium?  Take the nuclear modification factor $R_{\rm AA}$ of $J/\psi$ for example. Several models based on different physics assumptions do reproduce this quantity fairly well, making it difficult to pinpoint the relevant processes unambiguously. Therefore it has been proposed to investigate more discriminatory observables, such as the $\psi'$ to $J/\psi$ ratio, for which prediction by transport models or the statistical model of hadronization differ significantly. Here we wish to estimate this ratio using the in-medium spectra of fully equilibrated charmonium.

\begin{wrapfigure}{r}{0.5\textwidth}\vspace{-0.9cm}
  \begin{center}
   \includegraphics[scale=0.6]{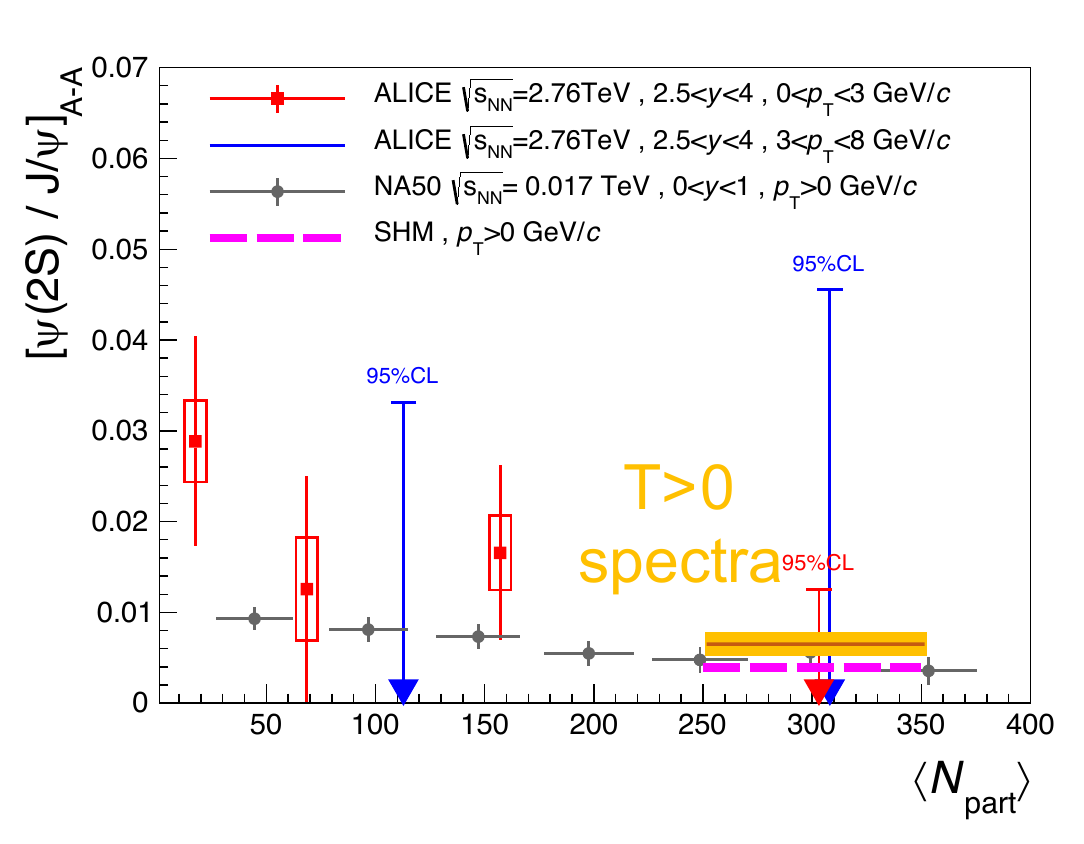}
  \end{center}\vspace{-0.9cm}
 \caption{Our estimate for the $\psi'/J/\psi$ ratio (orange) as well as that from the statistical model of hadronization (pink) and a recent first measurements by ALICE (adapted from \cite{Andronic:2015wma}).}\label{Fig:CompALICE}\vspace{-0.6cm}
\end{wrapfigure} 
The additional assumption we have to make here is that of an instantaneous freezeout, i.e.\ similar to the statistical model of hadronization, at $T_C=155$MeV we let all in-medium state go over their vacuum counterparts. We do so in the following way: the question we ask is how many vacuum states does the in-medium spectral peak in the charmonium spectral function correspond to? The answer we give in units of dilepton emission, which is related to the weighted area $R_{l\bar l}\propto \int dp_0 d^3{\bf p} \frac{\rho(P)}{P^2}n_B(p_0)$ under each spectral peak. I.e. we compute $R_{l\bar l}$ for both the in-medium and the $T=0$ spectrum and divide the two, leading us to an estimate of the number of states produced. If we do so for both $J/\psi$ and $\psi^\prime$ and divide the obtained values the outcome is
\begin{align}
\nonumber \left. \frac{N_{\psi'}}{ N_{J/\psi}} \right|_{T=T_C}=\frac{R_{\ell\bar\ell}^{\psi'}}{ R_{\ell\bar\ell}^{J/\psi}} \frac{ M_{\psi'}^2 |\Phi_{J/\psi}(0)|^2}{M_{J/\psi}^2 |\Phi_{\psi'}(0)|^2} =0.052\pm0.009, \,  \left.\frac{N_{\psi'}}{ N_{J/\psi}} \right|_{T=T_C} \frac{BR(\psi^\prime\to\mu^-\mu^+)}{BR(J/\psi\to\mu^-\mu^+)}=0.0069(11).
\end{align}
Compared to a recent first measurement of the ratio by the ALICE collaboration shown in Fig.~\ref{Fig:CompALICE}, we find that our result is slightly larger than the prediction by the statistical model of hadronization (SHM) but still well within the limits given by ALICE. While the difference to the SHM is larger than the estimated uncertainties it is not unfathomable that two approaches that are based on a full kinetic equilibration of the heavy quarks do show such a similar outcome.

\section{Conclusion}

The study of in-medium properties of heavy quarkonium from first principles has progressed significantly over the last years. The maturation of effective field theories, the arrival of realistic lattice simulations of QCD at finite temperature with almost physical pion mass and the development of novel Bayesian approaches to spectral function reconstruction have all contributed to further the computation of quarkonium in-medium spectral functions. Here we reported on two complementary approaches combining lattice QCD with the effective field theories NRQCD and pNRQCD. The former, direct approach, while describing realistic heavy-quarkonium with finite mass is still limited in resolution due to the small number of simulation datapoints available. The measured correlators indicated that overall in-medium modification proceeds hierarchically ordered according to the vacuum binding energy of the ground state in each channel. Using two different Bayesian spectral reconstructions (BR) and (MEM) we reconstructed the corresponding in-medium spectra and bracketed the melting temperatures of bottomonium S-wave and P-wave ground states from above and below. The systematic uncertainties of the different reconstruction methods were discussed.

The latter, indirect approach, as a first step required the computation of the in-medium heavy-quark potential, for which we showed our most recent results for full lattice QCD. Based on a generalized Gauss law in which the in-medium modification of  ${\rm Re}[V^{\rm QCD}]$ and ${\rm Im}[V^{\rm QCD}]$ is described by a single temperature dependent parameter $m_D$, we managed to describe our lattice data for the potential. Using the continuum corrected in-medium potential we then solved a Schr\"odinger equation to obtain the in-medium quarkonium spectra. They showed a hierarchical modification of the former vacuum states according to their binding energy. The peaks broaden and shift to lower masses, before vanishing into the continuum, which at the same time also moves to lower and lower energies. Using the criterion of the in-medium width being equal to the in-medium binding energy we defined and evaluated the melting temperatures of the individual states.

With the additional assumption of an instantaneous freezeout we estimated the $\psi^\prime$ to $J/\psi$ number ratio obtaining a value slightly larger than the estimate from the statistical model of hadronization.

The author thanks Y.~Burnier, O.~Kaczmarek, S.~Kim and P.~Petreczky for the fruitful collaboration in the presented studies. Calculations for \cite{Burnier:2015tda} were performed on the in-house cluster at the ITP in Heidelberg and the SuperB cluster at EPFL. This work is part of and supported by the DFG Collaborative Research Centre "SFB 1225 (ISOQUANT)".


\begin{thebibliography}{}
\bibitem{Andronic:2015wma} 
  A.~Andronic {\it et al.},
  Eur.\ Phys.\ J.\ C {\bf 76}, no. 3, 107 (2016)

\bibitem{Bazavov:2011nk} 
  A.~Bazavov {\it et al.},
  Phys.\ Rev.\ D {\bf 85}, 054503 (2012)
\bibitem{Borsanyi:2013bia} 
  S.~Borsanyi, Z.~Fodor, C.~Hoelbling, S.~D.~Katz, S.~Krieg and K.~K.~Szabo,
  Phys.\ Lett.\ B {\bf 730}, 99 (2014)

\bibitem{Bodwin:1994jh}
  G.~T.~Bodwin, E.~Braaten and G.~P.~Lepage,
  Phys.\ Rev.\ D {\bf 51} (1995) 1125

\bibitem{Brambilla:2004jw} 
  N.~Brambilla, A.~Pineda, J.~Soto and A.~Vairo,
  Rev.\ Mod.\ Phys.\  {\bf 77}, 1423 (2005) 


\bibitem{Thacker:1990bm}
  B.~A.~Thacker and G.~P.~Lepage,
  Phys.\ Rev.\ D {\bf 43} (1991) 196.

\bibitem{Lepage:1992tx}
  G.~P.~Lepage, L.~Magnea, C.~Nakhleh, U.~Magnea and K.~Hornbostel,
  Phys.\ Rev.\ D {\bf 46} (1992) 4052


\bibitem{Aarts:2010ek}
  G.~Aarts, S.~Kim, M.~P.~Lombardo, M.~B.~Oktay, S.~M.~Ryan,
  D.~K.~Sinclair, J.~-I.~Skullerud,
  Phys.\ Rev.\ Lett.\  {\bf 106 } (2011) 061602

\bibitem{Aarts:2011sm}
  G.~Aarts, C.~Allton, S.~Kim, M.~P.~Lombardo, M.~B.~Oktay, S.~M.~Ryan, D.~K.~Sinclair and J.~I.~Skullerud,
  `
  JHEP {\bf 1111} (2011) 103 

\bibitem{Aarts:2014cda}
  G.~Aarts, C.~Allton, T.~Harris, S.~Kim, M.~P.~Lombardo, S.~M.~Ryan and J.~I.~Skullerud,
  JHEP {\bf 1407} (2014) 097

\bibitem{Aarts:2013kaa}
  G.~Aarts, C.~Allton, S.~Kim, M.~Lombardo, S.~Ryan {\em et~al.},
  JHEP {\bf 1312} (2013) 064

\bibitem{Brambilla:2008cx}
  N.~Brambilla, J.~Ghiglieri, A.~Vairo and P.~Petreczky,
  Phys.\ Rev.\  D {\bf 78} (2008) 014017

\bibitem{Burnier:2007qm} 
  Y.~Burnier, M.~Laine and M.~Vepsalainen,
  JHEP {\bf 0801}, 043 (2008)

\bibitem{Jarrell:1996}
  J.~Skilling, S.F.~Gull,
  Lecture Notes-Monograph Series 20 (1991) 341;
  M.~Jarrell and J.E.~Gubernatis,
  Physics Reports, {\bf 269} (1996) 133

\bibitem{Asakawa:2000tr}
  M.~Asakawa, T.~Hatsuda, Y.~Nakahara,
  Prog.\ Part.\ Nucl.\ Phys.\  {\bf 46} (2001) 459

\bibitem{Burnier:2013nla}
  Y.~Burnier and A.~Rothkopf,
  Phys.\ Rev.\ Lett.\  {\bf 111} (2013) 18,  182003 
  
\bibitem{Kim:2014iga} 
  S.~Kim, P.~Petreczky and A.~Rothkopf,
  Phys.\ Rev.\ D {\bf 91}, 054511 (2015), Nucl.\ Phys.\ A {\bf 956}, 713 (2016)

\bibitem{Bazavov:2014pvz} 
  A.~Bazavov {\it et al.} [HotQCD Collaboration],
  Phys.\ Rev.\ D {\bf 90}, no. 9, 094503 (2014)


\bibitem{Laine:2006ns}
  M.~Laine, O.~Philipsen, P.~Romatschke and M.~Tassler,
  JHEP {\bf 0703} (2007) 054

\bibitem{Beraudo:2007ky}
  A.~Beraudo, J.~P.~Blaizot and C.~Ratti,
  Nucl.\ Phys.\  A {\bf 806} (2008) 312
  
\bibitem{Rothkopf:2009pk} 
  A.~Rothkopf, T.~Hatsuda and S.~Sasaki,
  PoS LAT {\bf 2009}, 162 (2009)
\bibitem{Rothkopf:2011db} 
  A.~Rothkopf, T.~Hatsuda and S.~Sasaki,
  Phys.\ Rev.\ Lett.\  {\bf 108}, 162001 (2012)
  

\bibitem{Burnier:2012az}
  Y.~Burnier and A.~Rothkopf,
  Phys.\ Rev.\ D {\bf 86} (2012) 051503,
  Phys.\ Rev.\ D {\bf 87}, 114019 (2013).
  
\bibitem{Burnier:2014ssa} 
  Y.~Burnier, O.~Kaczmarek and A.~Rothkopf,
  Phys.\ Rev.\ Lett.\  {\bf 114}, no. 8, 082001 (2015)

\bibitem{Burnier:2015nsa} 
  Y.~Burnier and A.~Rothkopf,
  Phys.\ Lett.\ B {\bf 753}, 232 (2016)


\bibitem{Burnier:2015tda} 
  Y.~Burnier, O.~Kaczmarek and A.~Rothkopf,
  JHEP {\bf 1512}, 101 (2015),

\bibitem{Burnier:2016kqm} 
  Y.~Burnier, O.~Kaczmarek and A.~Rothkopf,
  JHEP {\bf 1610}, 032 (2016).

\end{thebibliography}
\end{document}